\documentstyle[epsf]{article}
\begin{document}
\title{Gravitational vacuum polarization IV: \\
       Energy conditions in the Unruh vacuum}
\author{
Matt Visser${}^*$\\
Physics Department, Washington University, \\
Saint Louis, Missouri 63130-4899}
\date{gr-qc/9703001; 28 February 1997}
\maketitle

\begin{abstract}
Building on a series of earlier papers [gr-qc/9604007,9604008,9604009],
I investigate the various point-wise and averaged energy conditions
in the Unruh vacuum.  I consider the quantum stress-energy tensor
corresponding to a conformally coupled massless scalar field, work
in the test-field limit, restrict attention to the Schwarzschild
geometry, and invoke a mixture of analytical and numerical techniques.
I construct a semi-analytic model for the stress-energy tensor that
globally reproduces all known numerical results to within $0.8\%$,
and satisfies all known analytic features of the stress-energy
tensor. I show that in the Unruh vacuum (1) all standard point-wise
energy conditions are violated throughout the exterior region---all
the way from spatial infinity down to the event horizon, and (2)
the averaged null energy condition is violated on all outgoing
radial null geodesics. In a pair of appendices I indicate general
strategy for constructing semi-analytic models for the stress-energy
tensor in the Hartle--Hawking and Boulware states, and show that
the Page approximation is in a certain sense the minimal ansatz
compatible with general properties of the stress-energy in the
Hartle--Hawking state.
\end{abstract}

\bibliographystyle{unsrt}
\section{Introduction}

It is well-known that a quantum field theory constructed on a curved
background spacetime experiences gravitationally induced vacuum
polarization.  This effect typically induces a non-zero vacuum
expectation value for the stress-energy
tensor~\cite{deWitt75,Centenary,Birrell-Davies,Three-Hundred,Fulling,Visser}.
It is also well-known that this gravitationally induced vacuum
polarization will induce violations of at least some of the energy
conditions of classical general
relativity~\cite{Visser,Visser95,Visser96a,Visser96b,Visser96c}.

In earlier publications, I addressed these issue in some depth for
both the (3+1) Hartle--Hawking vacuum~\cite{Visser96a} and (3+1)
Boulware vacuum~\cite{Visser96b}, and also presented an analysis
for a (1+1)-dimensional toy model for which the exact analytic form
of the stress-energy tensor is known~\cite{Visser96c}.  For the
(3+1) Unruh vacuum there are additional subtleties that must be
addressed. The complications arise from two sources:
\begin{enumerate}
\item
For the (3+1) Unruh vacuum we do {\em not} have any approximate
analytic expression for the stress-energy tensor. Nothing along
the lines of the Page approximation for the (3+1) Hartle--Hawking
vacuum~\cite{Page82}, or the Page--Brown--Ottewill approximation
for the (3+1) Boulware vacuum~\cite{Brown-Ottewill}, has yet been
developed for the (3+1) Unruh vacuum.
\item
If we resort to numerical methods, using the numerical data of
either
Jensen--McLaughlin--Ottewill~\cite{JLO91,Jensen-private,Ottewill-private}
or Elster~\cite{Elster83}, we must be very careful to adequately
take into account a number of delicate and almost exact cancellations
between various components of the stress-energy tensor. Even though
individual components of the stress-energy tensor are often of
asymptotic order $O[(2GM/r)^2]$, the particular combination relevant
to the null energy condition [NEC] can easily be of asymptotic
order $O[(2GM/r)^3]$ or even smaller.
\end{enumerate}

To see where these cancellations come from, recall that the classic
analysis presented by Christensen and Fulling~\cite{Fulling77} is
enough to tell us that asymptotically at large $r$

\begin{equation}
\langle U | T^{\hat\mu\hat\nu} | U \rangle =  
{L\over4\pi r^2} \; \ell_+^{\hat\mu} \ell_+^{\hat\nu} + O[(2GM/r)^3].
\end{equation}

\noindent
Note that I choose to work in a local-Lorentz basis attached to
the fiducial static observers (FIDOS). The vector $\ell_+^{\hat\mu}
\equiv (1,1,0,0)$ is an outward-pointing radial null vector, $L$ is the
luminosity of the evaporating black hole, and the remaining terms
in the stress-energy fall off at least as rapidly as $1/r^3$.

If we look along the inward-pointing radial null vector
[$\ell_-^{\hat\mu} = (1,-1,0,0)$, that is: $\ell^-_{\hat\mu} =
(-1,-1,0,0)$], then

\begin{equation}
\langle U | T^{\hat\mu\hat\nu} | U \rangle  \; 
\ell^-_{\hat\mu} \ell^-_{\hat \nu} =  
{4 L\over4\pi r^2}+ O[(2GM/r)^3].
\end{equation}

\noindent
This is enough to tell you that at large enough distances the
Hawking flux will dominate, and that in this asymptotic region we
will therefore {\em satisfy} the NEC along inward-pointing null
geodesics.

On the other hand, if we look along the outward-pointing null
vector, then the term proportional to the luminosity ($L$) vanishes
and so at the very least

\begin{equation}
\langle U | T^{\hat\mu\hat\nu} | U \rangle  \; 
\ell^+_{\hat\mu} \ell^+_{\hat \nu}=  
O[(2GM/r)^3].
\end{equation}

\noindent
I shall actually prove something considerably stronger. I shall
prove that this quantity is negative everywhere outside the event
horizon and that

\begin{equation}
\langle U | T^{\hat\mu\hat\nu} | U \rangle  \; 
\ell^+_{\hat\mu} \ell^+_{\hat \nu}=  
O[(2GM/r)^5].
\end{equation}

\noindent
Thus for outward-pointing null geodesics the Hawking flux quietly
cancels, and it is the {\em sub-dominant} pieces of the stress-energy
that govern whether or not the NEC is satisfied.

To get around these difficulties, I have acquired the
Jensen--McLaughlin--Ottewill numerical
data~\cite{JLO91,Jensen-private,Ottewill-private} for the spin zero
Unruh vacuum, and after suitable refinements (to be more fully
described below), used this numerical data to construct a semi-analytic
three-parameter model that globally fits all the known numeric data
to better than $0.8\%$ accuracy.  Since the numerical data itself is
not expected to have better than $1\%$ accuracy, this is as good
as we can reasonably expect, and further refinements to the model
are not presently justifiable.

Once I have developed this semi-analytic model, all discussions of
the energy conditions in the Unruh vacuum will be phrased in terms
of this model, and I will not further use the numeric data itself.

As this work was nearing completion I became aware of a related
though distinct analysis by Matyjasek~\cite{Matyjasek}. I compare
and contrast my own analysis with that due to Matyjasek.

Finally, in two appendices, I sketch a general strategy for
constructing semi-analytic models for the Hartle--Hawking and
Boulware states.  I point out that there is a sense in which the
Page approximation is the minimal ansatz compatible with the known
behaviour of the stress-energy in the Hartle--Hawking state, and
show how to build a more general ansatz that does not disturb the
low-order terms coming from the Page approximation.

\section{Vacuum polarization in Schwarzschild spacetime}
\subsection{Covariant conservation of stress-energy}

By spherical symmetry we know that for {\em any} s-wave quantum
state $|\psi\rangle$

\begin{equation}
\langle \psi | T^{\hat\mu\hat\nu} | \psi \rangle \equiv 
\left[ \matrix{\rho&f&0&0\cr
               f&-\tau&0&0\cr
               0&0&p&0\cr
	       0&0&0&p\cr } \right].
\end{equation}

\noindent
Here $\rho$, $\tau$, $f$, and $p$ are functions of $r$, $M$, and
$\hbar$.  Note that I have now set $G\equiv1$, and continue to work
in a local-Lorentz basis attached to the fiducial static observers
(FIDOS). I shall work outside the horizon and will  make extensive
use of the dimensionless variable $z\equiv2M/r$.

In the classic paper by Christensen and Fulling~\cite{Fulling77},
it was shown that (by using the equations of covariant conservation)
the stress-energy tensor can, in the Schwarzschild spacetime, be
decomposed into four separately conserved quantities. These four
conserved tensors depend on two general functions of $r$ and two
integration constants. I choose to use a slightly different basis
for this decomposition, and write the stress-energy tensor as

\begin{equation}
\label{E-decomposition}
\langle \psi | T^{\hat\mu\hat\nu} | \psi \rangle \equiv 
 [T_{trace}]^{\hat\mu\hat\nu} +
 [T_{pressure}]^{\hat\mu\hat\nu} +
 [T_+]^{\hat\mu\hat\nu} +
 [T_-]^{\hat\mu\hat\nu}.
\end{equation}

\noindent
Here

\begin{equation}
 [T_{trace}]^{\hat\mu\hat\nu} \equiv
\left[ \matrix{-T(z)+{z^2\over1-z}H(z)&0&0&0\cr
               0&+{z^2\over1-z}H(z)&0&0\cr
               0&0&0&0\cr
	       0&0&0&0\cr } \right],
\end{equation}

\noindent
with

\begin{equation} 
\label{E-define-H}
H(z) \equiv {1\over2} \int_z^1{T(\bar z)\over {\bar z}^2} \; d{\bar z}. 
\end{equation}

\noindent
The conserved tensor $[T_{trace}]^{\hat\mu\hat\nu}$ depends only on the
trace $T$ of the total stress-energy tensor. Furthermore its trace
is equal to that of the full stress-energy tensor itself:
$[T_{trace}]^{\hat\mu}{}_{\hat\mu} = T$.

Next, I define
\begin{equation}
 [T_{pressure}]^{\hat\mu\hat\nu} =
\left[ \matrix{+2p(z)+{z^2\over1-z} G(z) &0&0&0\cr
               0&+{z^2\over1-z} G(z) &0&0\cr
               0&0&+p(z)&0\cr
	       0&0&0&+p(z)\cr } \right],
\end{equation}

\noindent
with

\begin{equation} 
\label{E-define-G}
G(z) \equiv 
\int_z^1 \left({2\over {\bar z}^3}-{3\over {\bar z}^2} \right) 
            \; p(\bar z) \; d{\bar z}.   
\end{equation}

The traceless conserved  tensor $[T_{pressure}]^{\hat\mu\hat\nu}$
depends only on the transverse pressure $p(z)$ of the total
stress-energy tensor.

This decomposition makes sense if the integrals $G(z)$ and $H(z)$
converge, which requires mild integrability constraints on $T(z)$
and $p(z)$ at the horizon. These constraints are certainly satisfied
for the Unruh (and Hartle--Hawking) vacuum state for which $T|_{z=1}$
and $p|_{z=1}$ are actually finite~\cite{Fulling77,JLO91,Elster83}.
Thus we can write

\begin{eqnarray}
\label{E-asymptotic-H}
H(z) &=& {1\over2} T|_{z=1} \; (1-z) + O[(1-z)^2],
\\
\label{E-asymptotic-G}
G(z) &=& - p|_{z=1}  \;(1-z) + O[(1-z)^2].
\end{eqnarray}

\noindent
This tells us that near the horizon

\begin{equation}
 [T_{trace}]^{\hat\mu\hat\nu}(z) \equiv
\left[ \matrix{-{1\over2}T|_{z=1}&0&0&0\cr
               0&+{1\over2}T|_{z=1}&0&0\cr
               0&0&0&0\cr
	       0&0&0&0\cr } \right] +O(1-z),
\end{equation}

\noindent
and

\begin{equation}
 [T_{pressure}]^{\hat\mu\hat\nu}(z) =
\left[ \matrix{+p|_{z=1}&0&0&0\cr
               0&-p|_{z=1}&0&0\cr
               0&0&+p|_{z=1}&0\cr
	       0&0&0&+p|_{z=1}\cr } \right] + O(1-z).
\end{equation}

\noindent
This is enough to imply that these two tensors are individually
regular at both the past and future horizons. (For the Boulware vacuum
minor changes in the formalism are required. See Appendix B.)

The function $G(z)$ can be somewhat rearranged by an integration
by parts. Suppose we define

\begin{equation}
\label{E-define-F}
F(z) \equiv 
\int_z^1 {\bar z}^2 \; (1-{\bar z}) 
\; {d\over d\bar z}\left({p(\bar z)\over{\bar z}^4}\right) \; d{\bar z}.
\end{equation}

\noindent
Then it is easy to show that 

\begin{equation}
G(z) = -{1-z\over z^2} p(z)  - F(z).
\end{equation}

\noindent
Near the horizon

\begin{equation}
\label{E-asymptotic-F}
F(z) = O[(1-z)^2].
\end{equation}

\noindent
Doing this again requires only the mild assumption that $p(z)$ is
well-behaved at the horizon $z=1$. Subject to this caveat the
tensor $[T_{pressure}]$ can be rewritten in the somewhat more
convenient form

\begin{equation}
 [T_{pressure}]^{\hat\mu\hat\nu} =
\left[ \matrix{+p(z)-{z^2\over1-z} F(z) &0&0&0\cr
               0&-p(z)-{z^2\over1-z} F(z) &0&0\cr
               0&0&+p(z)&0\cr
	       0&0&0&+p(z)\cr } \right].
\end{equation}

\noindent
In view of the definition of $F(z)$ this particular form of the
tensor makes it clear that a transverse pressure that falls of as
$O(z^4)$ will have special properties. 

Finally, I take

\begin{equation}
 [T_\pm]^{\hat\mu\hat\nu} =
  f_\pm \; \ell_{\pm}^{\hat\mu} \ell_{\pm}^{\hat\nu}=
  f_\pm \; {z^2\over 1-z} 
  \left[ \matrix{+1&\pm1&0&0\cr
                 \pm1&+1&0&0\cr
                  0&0&0&0\cr
	          0&0&0&0\cr } \right].
\end{equation}

\noindent
These two traceless conserved tensors correspond to out-going and
in-going null fluxes respectively. Furthermore $f_+$ and $f_-$ are
two {\em constants} with the dimensions of energy density that
determine the overall flux.  In terms of the overall flux $f(z)$
we have 

\begin{equation}
f(z) = (f_+ - f_-)\;  {z^2\over1-z}.
\end{equation}

\noindent
Note that $[T_+]$ is singular on the future horizon $H^+$, and
regular on the past horizon $H^-$. Conversely, $[T_-]$ is singular
on the past horizon $H^-$, and regular on the future horizon $H^+$.

It is easy to see that this decomposition is equivalent to that of
Christensen and Fulling. Perhaps the best starting point is equation
(2.5) on page 2090 of reference~\cite{Fulling77}. The functions
$H(z)$ and $G(z)$ defined above are {\em not} identical to those
of Christensen and Fulling, but are linear combinations of the
quantities appearing therein. Finally, the constants $K$ and $Q$
of that paper are in my basis given by $K= (f_+ - f_-) \, 4 M^4$, and
$Q= (-2 f_+) \, 4 M^4$, respectively. I have chosen this particular
basis because of its elegance, symmetry, and the ease with which
I can adapt it to my semi-analytic model to be introduced later in
this paper.

Everything done so far works for an arbitrary quantum field in an
arbitrary (spherically symmetric s-wave) quantum state in the
Schwarzschild geometry. (In particular it holds for both the Unruh and
Hartle--Hawking vacuum states, and with only minor modifications, also
for the Boulware vacuum state.) I will now start to particularize the
discussion.

\subsection{Conformally coupled quantum fields}

For any conformally coupled field, still in any arbitrary (spherically
symmetric) quantum state, the trace of the stress tensor is known
exactly and is given by the conformal anomaly. In a Schwarzschild
spacetime

\begin{equation}
T(z) = +\xi \; p_\infty \; z^6.
\end{equation}

\noindent
Here I have defined a constant (which in the Hartle--Hawking vacuum
can be interpreted as the pressure at spatial infinity) by

\begin{equation}
p_\infty \equiv {\hbar\over 90 (16\pi)^2 (2M)^4}. 
\end{equation}

\noindent
The number $\xi$ depends on the particular quantum field under
consideration. In particular, for a conformally coupled scalar
field $\xi=+96$.  For other conformally coupled quantum fields,
just change the $+96$ to the appropriate magic number. The relevant
coefficients can be easily deduced from the table on page 180 of
Birrell--Davies~\cite{Birrell-Davies} and are they are given in
Table I.

\begin{table}
\begin{center}
Table I. Anomaly coefficients.
\\[1ex]
\begin{tabular}{|c|c|c|c|c|} 
  \hline
  spin & Weyl anomaly & magic number & degeneracy & weight\\
  \hline
  $s$  & $a_2 = {1\over2880\pi^2}\{\#\,C^2 + \cdots \}$
                      & $T = \xi p_\infty z^6$&    $d$ &   $g$          \\
  \hline
  $0$         &   $-1$          &  $+96$     &  $1$   &    $1$         \\
  ${1\over2}$ &   $-{7\over4}$  &  $+168$    &  $2$   &    ${7\over4}$ \\
  $1$         &   $+13$         &  $-1248$   &  $2$   &    $2$         \\
  ${3\over2}$ &   $+{233\over4}$&  $-5592$   &  $2$   &    ${7\over4}$ \\
  $2$         &   $-212$        &  $+20352$  &  $2$   &    $2$         \\
  \hline
\end{tabular}
\\[1ex]
All fermions are non-chiral.
\end{center}
\end{table}

From the general definition (\ref{E-define-H}) above, the function
$H(z)$ is easily evaluated

\begin{equation} 
H(z) = {\xi\over10} \; p_\infty \; (1-z^5) = 
{\xi\over10} \; p_\infty \; (1-z) (1+z+z^2+z^3+z^4). 
\end{equation}

\noindent
Thus the trace contribution to the stress-energy is known exactly
and analytically, being given by a simple polynomial:

\begin{equation}
 [T_{trace}]^{\hat\mu\hat\nu} = {\xi\over10} \; p_\infty \; z^2 \;
\left[ \matrix{1+z+z^2+z^3-9z^4&0&0&0\cr
               0&1+z+z^2+z^3+z^4&0&0\cr
               0&0&0&0\cr
	       0&0&0&0\cr } \right].
\end{equation}

\noindent
In particular this implies that once we have a semi-analytic model
for the remaining free quantities $p(z)$, $f_+$, and $f_-$, no
matter how derived, we automatically have a model for the entire
stress-energy.

\subsection{Unruh vacuum}

In the Unruh vacuum, we have a lot of additional information
available beyond the rather general considerations given
above~\cite{Fulling77}. First, since the Unruh vacuum  is to be
regular on the future horizon we must have $f_+=0$, though it is
permissible to have $f_-\neq 0$. Naively, this appears to forbid
an outgoing flux, until we realize that there is nothing to stop
$f_-$ from being negative. In fact, this is exactly what happens,
and so I shall define a positive quantity $f_0$ and set $f_- = -
f_0$. This is why we often hear the assertion that (at the event
horizon) the Hawking radiation corresponds to an inward flux of
negative energy.

Collecting the terms in the general decomposition
(\ref{E-decomposition}) for the stress energy tensor, we have a this
stage

\begin{eqnarray}
\rho(z) &=& 
-f_0 {z^2\over1-z}
+{\xi\over10} \; p_\infty \; z^2 \left(1+z+z^2+z^3-9z^4\right)
\nonumber\\
&& 
\qquad +  p(z) - {z^2\over1-z} F(z).
\\
\tau(z) &=&
+f_0 {z^2\over1-z}
-{\xi\over10} \; p_\infty \; z^2  \left( 1+z+z^2+z^3+z^4 \right)
\nonumber\\
&&
\qquad +p(z) + {z^2\over1-z} F(z).
\\
f(z) &=&
+f_0 {z^2\over1-z}.
\end{eqnarray}

At asymptotic spatial infinity we want the stress-energy to look
like that of an outgoing flux of positive radiation~\cite{Fulling77}.
That is:  we need to have $\rho(z) \to f(z)$ asymptotically as $z
\to 0$.  From the Christensen--Fulling analysis, we know that
asymptotically the transverse pressure goes as $1/r^4$ [$O(z^4)$].
So, picking off the dominant terms [$O(z^2)$] at large distances
we see that

\begin{equation} 
{\xi\over 10} \; p_\infty -F(0) - f_0 = f_0. 
\end{equation}

\noindent
That is

\begin{equation}
f_0 = {\xi\over20} \; p_\infty - {F(0)\over2}.
\end{equation}

\noindent
Explicitly

\begin{eqnarray}
f_0 
&=& 
{\xi\over20} \; p_\infty - 
{1\over2}\int_0^\infty \bar z^2 (1-\bar z) 
{d\over d\bar z} \left( {p(\bar z)\over \bar z^4} \right) \; d\bar z 
\\
&=&
{\xi\over20} \; p_\infty + 
{1\over2}\int_0^1 \left({2\over {\bar z}^3}-{3\over {\bar z} ^2}\right) 
\; p(\bar z) \; d\bar z. 
\end{eqnarray}

\noindent
I shall have occasion to use this result as a critical internal test
for my semi-analytic model.

Substituting this result for $f_0$ back into the general expression
for the stress-energy, the various components are seen to be

\begin{eqnarray}
\rho(z) &=& 
-\left( {\xi\over20} \; p_\infty  - {F(0)\over2} \right) {z^2\over1-z}
+{\xi\over10} \; p_\infty \; z^2 \left(1+z+z^2+z^3-9z^4\right)
\nonumber\\
&& 
\qquad +  p(z) - {z^2\over1-z} F(z).
\\
\tau(z) &=&
+\left( {\xi\over20} \; p_\infty  - {F(0)\over2} \right) {z^2\over1-z}
-{\xi\over10} \; p_\infty \; z^2  \left( 1+z+z^2+z^3+z^4 \right)
\nonumber\\
&& 
\qquad +p(z) + {z^2\over1-z} F(z).
\\
f(z) &=&
+\left( {\xi\over20} \; p_\infty  - {F(0)\over2} \right) {z^2\over1-z}.
\end{eqnarray}

\subsection{Total luminosity: preliminary analysis}

The total luminosity of the black hole is now easily evaluated

\begin{equation} 
L = 4 \pi f_0 (2M)^2. 
\end{equation}
More explicitly
\begin{equation} 
L = 4 \pi (2M)^2 
\left[ {\xi\over20} \; p_\infty + 
       {1\over2}\int_0^1 
       \left( {2\over {\bar z}^3} - {3\over {\bar z} ^2} \right)
       \; p(\bar z) \; d\bar z \right].
\end{equation}

\noindent
This should be compared to the geometric optics approximation for
the luminosity, obtained from Stefan's law:

\begin{equation}
L_{geometric-optics} = {1\over2} \; \sigma \; T^4 \; S.
\end{equation}

\noindent
The factor $1/2$ comes from the fact that there is only one
polarization state for a scalar field, while Stefan's constant is, in
the current geometrodynamic units, simply $\sigma = \pi^2 /60$.  [For
SI purists: $\sigma = (\pi^2/60) (k^4/\hbar^3 c^2)$.] For the
temperature we take the Hawking temperature $T = T_H = 1/(8\pi M)$.
[For SI purists: $k T_H = (\hbar c^2)/(8\pi G M)$.] Finally, the
effective radiating surface area is $S= 4 \pi (3\sqrt{3} M)^2 = 108
\pi M^2$. [For SI purists: $S = 108 \pi (GM/c^2)^2$.] The effective
surface area is deduced from the well-known textbook result that any
null geodesic of impact parameter less than $a=3\sqrt{3} M$ will fall
into the black hole~\cite{MTW,Weinberg,Wald}.  Equivalently, the
geometric optics absorption cross section for null particles is
$\sigma_{cross-section} = 27 \pi M^2$~\cite{Fulling77}.  Pulling this
all together

\begin{equation}
L_{geometric-optics} 
= 81 \pi \; M^2 \; p_\infty 
= {9\over 40960 \pi M^2}
= 6.99411\times 10^{-5} M^{-2}. 
\end{equation}

\noindent
Restated as an estimate of the flux, we have
\begin{equation}
f_{geometric-optics} 
= {81\over 16} \; p_\infty.
\end{equation}

\subsection{Pointwise energy conditions: preliminary analysis}

From the analytic discussion presented so far, we have enough
information to make a preliminary analysis of the pointwise energy
conditions. I shall extract a very simple and robust result using
a minimum amount of analytic information.

We have already seen in the introduction that if we look along
the outward-pointing radial null direction things simplify
dramatically. So let's use what we have so far and calculate

\begin{eqnarray}
\label{E-nec-preliminary}
\langle U | T^{\hat\mu\hat\nu} | U \rangle  \; 
\ell^+_{\hat\mu} \ell^+_{\hat \nu} 
&=&   
\rho -\tau - 2 f, 
\\
&=&
-4 f_0 {z^2\over1-z} 
+ {\xi\over5}\; p_\infty z^2(1+z+z^2+z^3-4z^4) 
\nonumber\\
&& 
\qquad - 2 {z^2\over1-z} F(z),
\\
&=&
-4 f_0 {z^2\over1-z} 
+{\xi\over5}\; p_\infty \; z^2(1-z)(1+2z+3z^2+4z^3) 
\nonumber\\
&& 
\qquad - 2 {z^2\over1-z} F(z).
\end{eqnarray}

\noindent
If we now work close to the horizon, and use the near-horizon
asymptotic form for $F(z)=O[(1-z)^2]$ as deduced in eq.
(\ref{E-asymptotic-F}), then

\begin{eqnarray}
\langle U | T^{\hat\mu\hat\nu} | U \rangle \; 
\ell^+_{\hat\mu} \ell^+_{\hat \nu} 
&=&   
- {4 f_0\over1-z} + 8 f_0 + O(1-z).
\end{eqnarray}

\noindent
So, provided only that the luminosity is positive, we will definitely
have violations of the null energy condition sufficiently near the
event horizon. This automatically implies that the weak, strong,
and dominant energy conditions will also be violated sufficiently
close to the event horizon.

Notice how minimal the input to this result has been: If the black
hole radiates, and the conserved stress-energy has the correct
asymptotic behaviour for the Unruh vacuum (both near the horizon
and at spatial infinity), then the null energy condition must be
violated sufficiently near the event horizon.

This is in complete agreement with the celebrated area increase
theorem of black hole dynamics~\cite{Hawking-Ellis}. From the area
increase theorem we know that if a black hole radiates, {\em and
if that radiation back-reacts on the black hole so as to decrease
its total mass}, then the null energy condition must be violated
somewhere in the spacetime. The area increase theorem is of course
derived using totally different techniques.

Naturally, once I have finished setting up my semi-analytic model,
I will be saying a whole lot more about violations of the various
energy conditions. In fact, I shall show that these violations of
the point-wise energy conditions persist all the way from the
horizon out to spatial infinity.

\subsection{Model building: preliminary analysis}

Now we know, both from the analytical analysis of
Christensen--Fulling~\cite{Fulling77}, and the numerical analyses
of Jensen--McLaughlin--Ottewill~\cite{JLO91} and Elster~\cite{Elster83},
that the transverse pressure is well-behaved all the way down to
the horizon. This suggests that the transverse pressure might be
usefully modelled by some simple convergent power series in $z=2M/r$.
We also know that at large distances the transverse pressure must
fall as $z^4$. Suppose we normalize to $p_\infty$, introduce a
denumerably infinite set of dimensionless coefficients $k_n$,  and
define

\begin{equation}  
p(z) \equiv  p_\infty \; \sum_{n=4}^\infty k_n z^n. 
\end{equation}

\noindent
Then, from the definition (\ref{E-define-F}) of $F(z)$

\begin{eqnarray}
F(z) 
&=&  p_\infty \sum_{n=5}^\infty (n-4) k_n 
     \left[ {1-z^{n-2}\over n-2} - {1-z^{n-1}\over n-1 } \right]
\\
&=&  p_\infty \sum_{m=3}^\infty  {1-z^m\over m} 
     \left[  (m-2) k_{m+2} - (m-3) k_{m+1} \right].
\end{eqnarray}

\noindent
In particular

\begin{equation}
F(0) =  p_\infty \sum_{n=5}^\infty {(n-4) k_n \over (n-2)(n-1) }.
\end{equation}

\noindent
Thus

\begin{equation}
f_0 = p_\infty 
\left( {\xi\over20} - \sum_{n=5}^\infty {(n-4) k_n\over2(n-2)(n-1)} \right)
\end{equation}

\noindent
A brief calculation yields several equivalent forms for $F(z)$

\begin{eqnarray}
F(z) 
&=&  p_\infty \left\{  
      \sum_{n=5}^\infty {(n-4) k_n\over(n-2)(n-1)} -  
      \sum_{n=5}^\infty (n-4) k_n
              \left[ {z^{n-2}\over n-2} - {z^{n-1}\over n-1 } \right] 
              \right\}
\label{E-O3}
\\
&=&  p_\infty (1-z) 
\Bigg\{ \sum_{n=5}^\infty {(n-4) k_n\over(n-2)(n-1)} \times
\nonumber\\
&& 
\qquad \left[ 1 + z + z^2 + \cdots + z^{n-3} - (n-2)z^{n-2} \right] \Bigg\}
\label{E-diff-F}
\\
&=&  p_\infty (1-z) \Bigg\{ \sum_{m=3}^\infty 
     \left[  {(m-2) k_{m+2} - (m-3) k_{m+1}\over m}  \right] \times
\nonumber\\
&& 
\qquad \left[ 1 + z + z^2 + \cdots + z^{m-1} \right] \Bigg\}
\\
&=& p_\infty (1-z)^2  
\Bigg\{ 
  \sum_{n=5}^\infty {(n-4) k_n\over(n-2)(n-1)} \times
\nonumber\\
&& 
\qquad  \left[ 1 + 2z + 3 z^2 + \cdots +z^{n-4}+ (n-2)z^{n-3} \right] 
\Bigg\}
\label{E-factor-2}
\end{eqnarray}

\noindent
The multiple expressions given above are useful in different
situations. The first expression is most useful near spatial infinity
(where it implies $F(z) = F(0) + O(z^3)$\label{D-O3} and makes
manifest the rapid falloff of the remainder of the stress-energy
once the Hawking flux has been subtracted off).  The second, third,
and fourth expressions are most useful near the horizon, where the
explicit $(1-z)$ factor helps to keep the behaviour of this particular
term mild. Finally, the third form given above can also be cast
into the form

\begin{equation}
F(z) =
  p_\infty (1-z) \sum_{\ell=0}^\infty z^\ell 
     \left[ \sum_{j=max(\ell+1,3)}^\infty
       {(j-2) k_{j+2} - (j-3) k_{j+1} \over j} 
     \right].
\end{equation}

If we work near spatial infinity this is enough to show that

\begin{eqnarray}
\rho(z) &=& 
+\left[{\xi\over20} \; p_\infty - {F(0)\over2}\right] { z^2\over1-z}
+ O(z^4). 
\\
\tau(z) &=&
-\left[{\xi\over20} \; p_\infty - {F(0)\over2}\right] { z^2\over1-z}
+ O(z^4). 
\\
p(z) &=& O(z^4).
\\
f(z) &=&
+\left[ {\xi\over20} \; p_\infty  - {F(0)\over2} \right] {z^2\over1-z}.
\end{eqnarray}

\noindent
On the other hand, near the horizon we have

\begin{eqnarray}
\rho(z) &=& 
-\left[{\xi\over20} \; p_\infty - {F(0)\over2}\right] { z^2\over1-z}
+ O(1). 
\\
\tau(z) &=&
+\left[{\xi\over20} \; p_\infty - {F(0)\over2}\right] { z^2\over1-z}
+ O(1). 
\\
p(z) &=& O(1).
\\
f(z) &=&
+\left[ {\xi\over20} \; p_\infty  - {F(0)\over2} \right] {z^2\over1-z}.
\end{eqnarray}

\noindent
Note the crucial change in the signs of the energy density and radial
tension needed to keep the stress-energy regular on the future
horizon.

The second form of $F(z)$ given above can now be substituted into the
stress-energy to show

\begin{eqnarray}
\rho(z) &=& 
-\left[ 
 {\xi\over20} \; p_\infty  - \sum_{n=5}^\infty {(n-4) k_n\over2(n-2)(n-1)}
 \right] {z^2\over1-z}
\nonumber\\
&& 
+{\xi\over10} \; p_\infty \; z^2 \left(1+z+z^2+z^3-9z^4\right)
+  p_\infty \sum_{n=4}^\infty k_n z^n
\nonumber\\
&& 
 -  p_\infty z^2  \sum_{n=5}^\infty {(n-4) k_n\over(n-2)(n-1)} 
     \left[ 1 + z + z^2 + \cdots + z^{n-3} - (n-2)z^{n-2} \right].
\nonumber\\
&&
\\
\tau(z) &=&
+\left[ 
 {\xi\over20} \; p_\infty  - \sum_{n=5}^\infty {(n-4) k_n\over2(n-2)(n-1)} 
\right] {z^2\over1-z}
\nonumber\\
&& 
-{\xi\over10} \; p_\infty \; z^2  \left( 1+z+z^2+z^3+z^4 \right)
+  p_\infty \sum_{n=4}^\infty k_n z^n
\nonumber\\
&& 
 +  p_\infty z^2 \sum_{n=5}^\infty {(n-4) k_n\over(n-2)(n-1)} 
     \left[ 1 + z + z^2 + \cdots + z^{n-3} - (n-2)z^{n-2} \right].
\nonumber\\
&&
\\
f(z) &=&
+\left[ 
 {\xi\over20} \; p_\infty  - \sum_{n=5}^\infty {(n-4) k_n\over2(n-2)(n-1)} 
\right] {z^2\over1-z}.
\end{eqnarray}

\noindent
From the preceding analysis it is clear that this model satisfies all
the known properties of the stress-energy tensor in the Unruh vacuum
(anomalous trace, covariant conservation, asymptotic behaviour both at
spatial infinity and the horizon), and is the most general form of the
stress-energy to do so. Consequently these equations provide a general
formalism for the stress-energy tensor in the Unruh vacuum.

\subsection{Model building: the final model}

With these analytic preliminaries out of the way, the final construction
of the semi-analytic model is quite anti-climactic: Merely truncate
the power series for $p(z)$ at some finite integer $n$ to obtain
a rational polynomial approximation to the stress-energy tensor.

We know that the power series starts off at order $z^4$. Further
we know that the anomalous trace introduces order $z^6$ terms. So
good first guess is to simply try the three-term polynomial

\begin{equation}
p(z) = p_\infty \; z^4 \; ( k_4 + k_5 z + k_6 z^2 ).
\end{equation}

\noindent
Remarkably, this ansatz is sufficient to fit all the known numeric
data within expected tolerances.  Substituting this ansatz into
the formulae above completes the model:

\begin{eqnarray}
\rho(z) &=& 
-\left( 
 {\xi\over20} - {k_5\over24} - {k_6\over20}
 \right)  \; p_\infty \; {z^2\over1-z}
+{\xi\over10} \; p_\infty \; z^2 \left(1+z+z^2+z^3-9z^4\right)
\nonumber\\
&&
+  p_\infty ( k_4 z^4 + k_5 z^5 + k_6 z^6 )
\nonumber\\
&& 
-  p_\infty z^2   
     \left[ {k_5\over12} (1 + z + z^2 -3z^3) + 
            {k_6\over10} (1+z+z^2+z^3-4z^4) 
      \right].
\\
\tau(z) &=&
+\left( 
 {\xi\over20} - {k_5\over24} - {k_6\over20}
\right)  \; p_\infty \; {z^2\over1-z}
-{\xi\over10} \; p_\infty \; z^2  \left( 1+z+z^2+z^3+z^4 \right)
\nonumber\\
&&
+  p_\infty ( k_4 z^4 + k_5 z^5 + k_6 z^6 )
\nonumber\\
&& 
+  p_\infty z^2 
     \left[ {k_5\over12} (1 + z + z^2 -3z^3) + 
            {k_6\over10} (1+z+z^2+z^3-4z^4) 
      \right].
\\
f(z) &=&
+\left( 
 {\xi\over20}  - {k_5\over24} - {k_6\over20}
\right)  \; p_\infty \; {z^2\over1-z}.
\end{eqnarray}

\noindent
A little brute force yields

\begin{eqnarray}
\rho &=& {z^2\over1-z} p_\infty
  \Bigg[ \left({\xi\over20} - {k_5\over24} - {k_6\over20}\right) + 
          k_4 z^2 + 
          \left(-k_4 + {4 k_5\over3}\right) z^3 
\nonumber\\
&&\qquad 
       + \left({-\xi} -{5 k_5\over4} + {3 k_6\over2} \right)z^4 
       + \left( {9 \xi\over10} -{7 k_6\over5} \right)z^5 \Bigg].
\\
\tau &=& {z^2\over1-z} p_\infty
  \Bigg[ - \left({\xi\over20} - {k_5\over24} - {k_6\over20}\right) 
         + k_4  z^2 + 
          \left(-k_4 + {2 k_5\over3}\right) z^3 
\nonumber\\
&&\qquad 
       + \left(-{3 k_5\over4} + {k_6\over2} \right)z^4 
       + \left( {\xi\over10} -{3 k_6\over5} \right)z^5 \Bigg].
\end{eqnarray}

\noindent
Performing a least squares fit on the transverse pressure data yields

\begin{equation}
p(z) = p_\infty z^4 ( 26.5652  - 59.0214 z  + 38.2068 z^2 ).
\end{equation}

\noindent
The energy density and radial tension are then derived quantities
with the values

\begin{eqnarray}
\rho &=& {z^2\over1-z}
\left(
5.34889  + 26.5652 z^2  - 105.260 z^3  + 35.0869 z^4  + 32.9105 z^5
\right).
\nonumber\\
&&
\\
\tau &=& {z^2\over1-z} 
\left( 
-5.34889 + 26.5652 z^2  - 65.9128 z^3  + 63.3694 z^4  - 13.3241 z^5
\right).
\nonumber\\
&&
\end{eqnarray}

\noindent 
(I am quoting six figure accuracy only for clarity of exposition---we
should not trust these formulae beyond the 1\% level.)

This data-fitting was performed in the following manner: Through
the kind offices of Professor Ottewill and Professor Jensen I
acquired copies of the numerical data they used in their 1991
paper~\cite{JLO91,Jensen-private,Ottewill-private}. For spin zero,
they provided me with the expectation values of the stress-energy
tensor in the Unruh vacuum, which they had calculated by numerically
evaluating the difference, $\langle U | p(z) | U \rangle - \langle
H | p(z) | H \rangle $, between the transverse pressures in the
Unruh and Hartle--Hawking states, and adding this difference to
the previously published data of Howard~\cite{Howard}. After checking
their data for internal consistency~\cite{JLO-consistency}, I
subtracted Howard's values for $\langle H | p(z) | H \rangle$ in
order to reconstruct these differences. I then added these differences
to the improved Anderson--Hiscock--Samuel values for  $\langle H
| p(z) | H \rangle$~\cite{AHS,Anderson-private}. I discarded the
point on the horizon, since the Hartle--Hawking data at the horizon
is itself an extrapolation. Also, because the various data sets
are calculated for somewhat different values of $r$, I had to
discard several other data points when merging the sets. Finally
I added the point at spatial infinity because we have exact
information at that point.  The resulting data set contained 26
points, and is summarized in the table given herein. (To avoid
unnecessary proliferation of numerical data, I am including only
the bare minimum required to construct the semi-analytic model.)
Mathematica was then used to perform a simple least-squares fit,
thereby producing the coefficients given above.

By comparing the fitted curve to the original data set, I checked that the
maximum deviation was $0.8\%$. Given the expected $1\%$ accuracy
of the numeric difference data further refinement of this model
does not seem currently justifiable. When graphically plotted the
actual numerical data cannot be visually distinguished from the
fit.

\begin{table}[htbp]
\begin{center}
Table II. Numerical data used to create the semi-analytic model.
\\[1ex]
\begin{tabular}{|c|c|c|c|}
  \hline
  $r/2M$    &   $p_H$   &$p_U - p_H$&   $p_U$   \\
  \hline
  1.05 &       8.54318 &  -4.426 &     4.11718 \\
  1.10 &       7.37818 &  -4.312 &     3.06618 \\
  1.15 &       6.56051 &  -4.194 &     2.36651 \\
  1.20 &       5.96614 &  -4.076 &     1.89014 \\
  1.25 &       5.51801 &  -3.962 &     1.55601 \\
  1.30 &       5.16771 &  -3.848 &     1.31971 \\
  1.35 &       4.88431 &  -3.740 &     1.14431 \\
  1.40 &       4.64780 &  -3.634 &     1.01380 \\
  1.45 &       4.44508 &  -3.536 &     0.90908 \\
  1.50 &       4.26740 &  -3.440 &     0.82740 \\
  1.55 &       4.10898 &  -3.350 &     0.75898 \\
  1.60 &       3.96552 &  -3.264 &     0.70152 \\
  1.65 &       3.83431 &  -3.182 &     0.65231 \\
  1.70 &       3.71331 &  -3.108 &     0.60531 \\
  1.75 &       3.60103 &  -3.034 &     0.56703 \\
  1.80 &       3.49635 &  -2.966 &     0.53035 \\
  1.85 &       3.39840 &  -2.900 &     0.49840 \\
  1.90 &       3.30648 &  -2.838 &     0.46848 \\
  1.95 &       3.22005 &  -2.780 &     0.44005 \\
  2.00 &       3.13862 &  -2.724 &     0.41462 \\
  2.10 &       2.98924 &  -2.620 &     0.36924 \\
  2.20 &       2.85568 &  -2.528 &     0.32768 \\
  2.30 &       2.73583 &  -2.444 &     0.29183 \\
  2.40 &       2.62795 &  -2.368 &     0.25995 \\
  2.50 &       2.53057 &  -2.298 &     0.23257 \\
 Infinity&     1.00000 &  -1.000 &     0.00000 \\
  \hline
\end{tabular}
\\[1ex]
The Hartle--Hawking data is from
Anderson--Hiscock--Samuel~\cite{AHS,Anderson-private}. 
\\
The difference data is inferred from
Jensen--McLaughlin--Ottewill~\cite{JLO91,Jensen-private,Ottewill-private}.
\\
The final column is the input to the least-squares fit.
\end{center}
\end{table}

Once a fit to $p(z)$ has been agreed upon, there is no further
flexibility in the model, and $\rho(z)$, $\tau(z)$, and the flux
$f(z)$ [equivalently, the total luminosity $L$] are completely
specified. In particular, it is meaningless to try independent
curve fitting to the energy density or radial tension.

Indeed, if we obtain a good fit to the transverse pressure, which
then does {\em not} result in a good fit to the energy density or
radial tension, this does not mean that the fit is bad. On the
contrary, since the transverse pressure is used (in both the
semi-analytic model and the numerical estimates) to {\em calculate}
the energy density and radial tension any discrepancy in these
quantities must be ascribed to actual error (either numerical
roundoff of something more systematic) rather than failure of the
fit~\cite{JLO-energy-density}.

\begin{figure}[htbp]
\vbox{\hfil\epsfbox{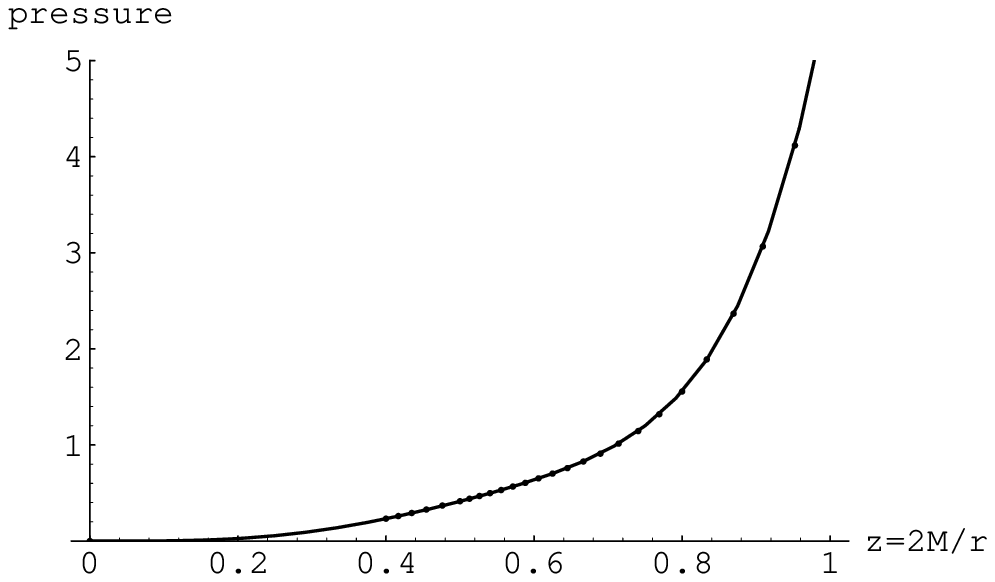}\hfil}
\caption[Transverse pressure.]
{\label{fig1}
This graph shows the transverse pressure in the Unruh vacuum, both
fitted and numerically estimated, as a function of $z=2M/r$ over
the entire available data range.}
\end{figure}

\begin{figure}[htbp]
\vbox{\hfil\epsfbox{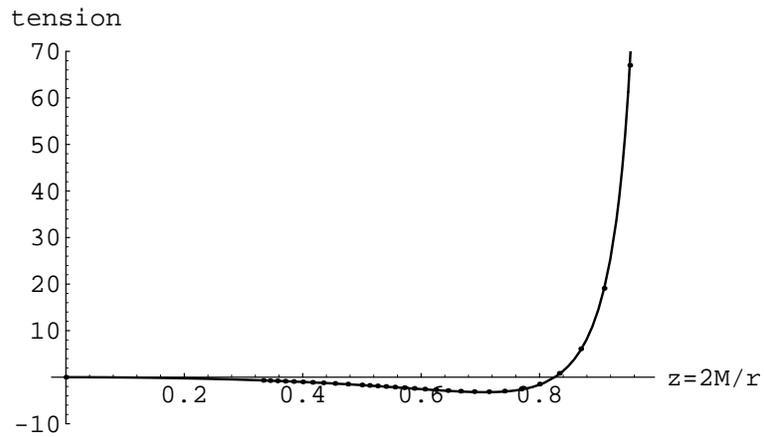}\hfil}
\caption[Radial tension.]
{\label{fig2}
This graph shows the radial tension in the Unruh vacuum, both fitted
and numerically estimated, as a function of $z=2M/r$, over the
entire available data range.}
\end{figure}

\begin{figure}[htbp]
\vbox{\hfil\epsfbox{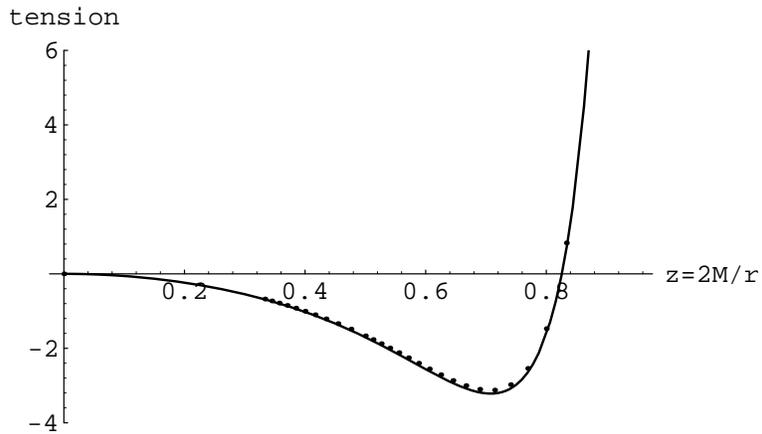}\hfil}
\caption[Radial tension.]
{\label{fig3}
This graph shows the radial tension in the Unruh vacuum, both fitted
and numerically estimated, as a function of $z=2M/r$, focussing on
small $z$.}
\end{figure}

\begin{figure}[htbp]
\vbox{\hfil\epsfbox{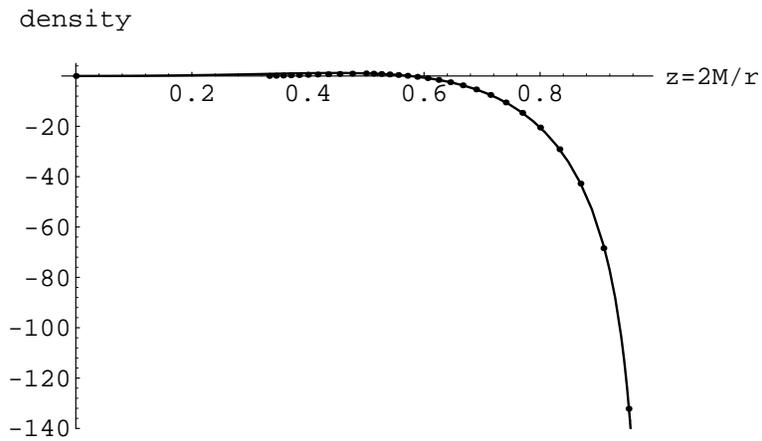}\hfil}
\caption[Energy density.]
{\label{fig4}
This graph shows the energy density in the Unruh vacuum, both fitted
and numerically estimated, as a function of $z=2M/r$, over the
entire available data range.}
\end{figure}

\begin{figure}[htbp]
\vbox{\hfil\epsfbox{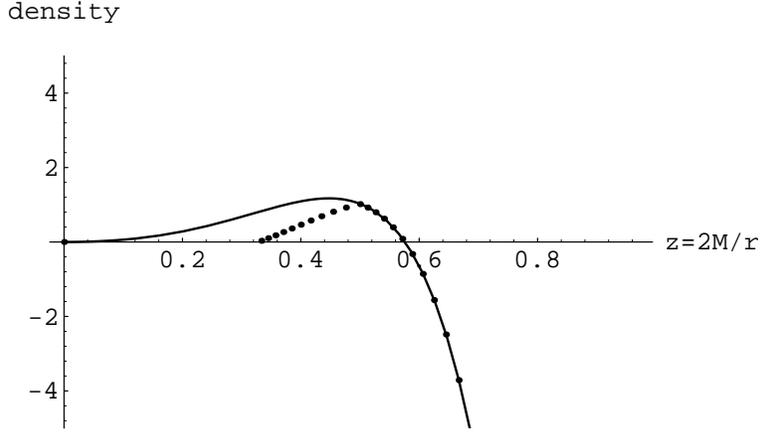}\hfil}
\caption[Energy density.]
{\label{fig5}
This graph shows the energy density in the Unruh vacuum, both fitted
and numerically estimated, as a function of $z=2M/r$, focussing on
small $z$. Note the discrepancy in the region below $z=0.5$, $r>4M$.
(See footnote~\cite{JLO-energy-density}.)
}
\end{figure}

There is an important consistency check that the semi-analytic
model should pass: Elster has calculated~\cite{Elster83} the total
luminosity of a black hole against scalar emission using standard
techniques~\cite{Page76} which are totally orthogonal to the present
analysis.  At each frequency he calculates a transmission coefficient
which describes how much of the radiation leaving the event horizon
actually makes it out to spatial infinity. [In (3+1) dimensions
some radiation is always back-scattered from the gravitational
field.] After multiplying this transmission coefficient by a
Planckian spectrum and integrating over all frequencies Elster
reports the total scalar luminosity as

\begin{equation}
L_{Elster} = 7.44\times10^{-5} M^{-2}.
\end{equation} 

\noindent
In the notation of this paper, this translates to

\begin{equation}
f_{Elster} = {81\over16} {7.44\over6.9941} \; p_\infty = 5.385 \; p_\infty.
\end{equation}

\noindent
On the other hand, the semi-analytic model gives

\begin{equation}
f_0 
= 
\left({96\over20}+ {59.0214\over24} - {38.2068\over20} \right) \; p_\infty 
=
5.349 \;  p_\infty
\end{equation}

\noindent
That two such rather different techniques give total luminosities
in such good agreement (better that $0.7\%$) is not only very
encouraging, but is in fact a powerful consistency check on the
numeric data. This completes construction of the semi-analytic
model.

\subsection{Matyjasek's analysis}

To compare my model to the model developed by Matyjasek~\cite{Matyjasek}
we should first note that Matyjasek does not base his model on the
general Christensen--Fulling analysis as outlined and developed
above, but instead uses a less general ansatz based on the
Brown--Ottewill approximation~\cite{Brown-Ottewill}. What he calls
his $N=6$ ansatz is (as presented) internally inconsistent. It is
incapable of simultaneously fitting the luminosity data and giving
the right asymptotic behaviour at infinity.

More precisely, if we fix the correct asymptotic behaviour at
spatial infinity, then the luminosity is not a free parameter in
Matyjasek's $N=6$ ansatz. His ansatz is equivalent to keeping $k_4$
a free variable and setting $k_5=+6$ and $k_6=-9$, in which case
$F(0) = -2/5$ and $f_0 = 5 p_\infty$. (This is close to but not
equal to the geometric optics approximation $f_{geometric-optics}
\equiv 81/16=5.0625\neq5$; it is certainly not equal to Elster's
calculated value $f_{Elster} = 5.385$.) This analysis also forces
Matyjasek's $b_2$ coefficient to be exactly $b_2=5$.  (Matyjasek's
$T$ is my $p_\infty$.) With only one free parameter ($b_4$) the
$N=6$ ansatz can at best give only qualitative agreement with the
numeric data.

Matyjasek's $N=7$ ansatz has enough free parameters to fit both
the asymptotic behaviour of the stress-energy and leave the luminosity
as a free variable. This model is equivalent to setting $k_6=-9$,
$k_5=126-24f_0/p_\infty$, and $k_4 = [p|_{z=1} +24f_0]/p_\infty -
117$. So in this model $F(0) = -2f_0 +(96/10)p_\infty$. Matyjasek's
$N=7$ model has only two free parameters, $f_0$ and $p|_{z=1}$,
the flux and the transverse pressure on the horizon, which he fits
to these two single pieces of data as calculated by Elster and
Jensen--McLaughlin--Ottewill. Matyjasek's analysis does not use
any of the other numeric data in any quantitative manner. With only
two free parameters the $N=7$ ansatz still  gives only qualitative
agreement with the numeric data.

In contrast, my model has three completely free parameters, $k_4$,
$k_5$, and $k_6$. I perform a global unconstrained fit to the
totality of the available data, and provide a quantitative statement
on the quality of the fit ($1\%$ accuracy). Furthermore, I use
Elster's luminosity calculation as a consistency check rather than
as input.

\section{Energy conditions}

Outside the event horizon, the null energy condition (NEC; $\langle
T_{\hat\mu\hat\nu} k^{\hat\mu} k^{\hat\nu} \geq 0?$) can be completely
analyzed by looking at three special cases: outgoing null vectors,
ingoing null vectors, and transverse null vectors. The NEC reduces to
the three constraints

\begin{equation}
\rho(r) - \tau(r) \pm 2 f\geq 0? \qquad \rho(r) + p(r) \geq 0?
\end{equation}

\subsection{Numerical analysis using the semi-analytic model}

Numerically, these conditions are best investigated visually, by
inspection of the relevant graphs.

\begin{figure}[htbp]
\vbox{\hfil\epsfbox{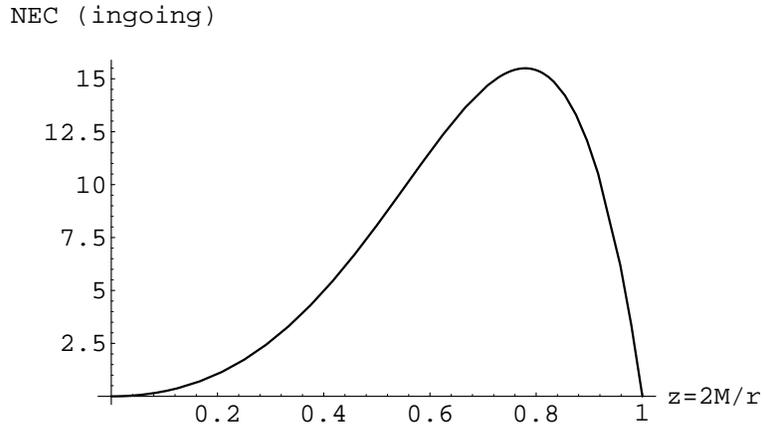}\hfil}
\caption[NEC (ingoing)]
{\label{fig6}
$\rho-\tau+2f$: This graph shows the NEC (evaluated using the
semi-analytic model) on ingoing radial null vectors as a function
of $z=2M/r$. Note that the curve is everywhere positive, so that
the NEC is {\em satisfied} on ingoing null geodesics.}
\end{figure}

\begin{figure}[htbp]
\vbox{\hfil\epsfbox{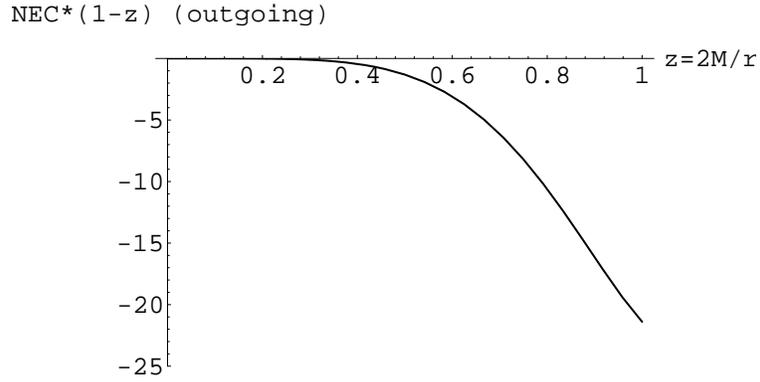}\hfil}
\caption[NEC (outgoing)]
{\label{fig7}
$\rho-\tau-2f$: This graph shows the NEC (evaluated using the
semi-analytic model and multiplied by $(1-z)$ to control the singularity
at the horizon) on outgoing radial null vectors as a function of
$z=2M/r$. Note that the curve is everywhere negative, so that the
NEC is {\em violated} on ingoing null geodesics.}
\end{figure}

\begin{figure}[htbp]
\vbox{\hfil\epsfbox{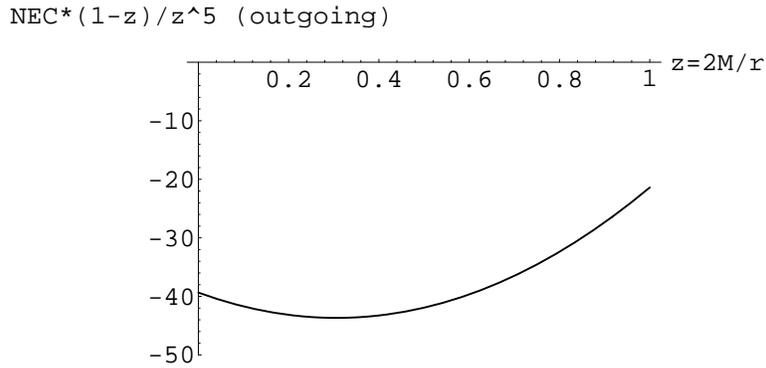}\hfil}
\caption[NEC (outgoing)]
{\label{fig8}
$\rho-\tau-2f$: This graph shows the NEC (evaluated using the
semi-analytic model, multiplied by $(1-z)$ to control the singularity
at the horizon, and divided by $z^5$ to suppress the asymptotic
behaviour at infinity) on outgoing radial null vectors as a function
of $z=2M/r$. Note that the curve is everywhere negative, so that
the NEC is {\em violated} on outgoing null geodesics.  }
\end{figure}

\begin{figure}[htbp]
\vbox{\hfil\epsfbox{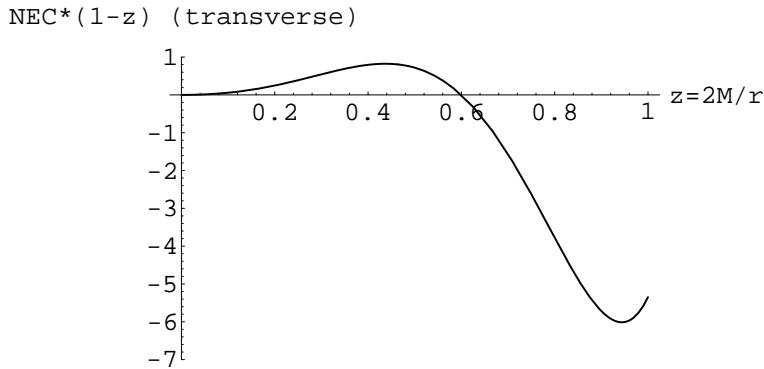}\hfil}
\caption[NEC (transverse)]
{\label{fig9} $\rho+p$: This graph shows the NEC (evaluated using
the semi-analytic model and multiplied by $(1-z)$ to control the
singularity at the horizon) on transverse null vectors as a function
of $z=2M/r$. Note that the curve changes sign near $z=0.6$.  The
NEC is {\em satisfied} on transverse null vectors far away from
the black hole, and {\em violated} sufficiently near the black
hole. (In particular the NEC is {\em violated} on the unstable
photon orbit at $r=3M$.)}
\end{figure}

\begin{figure}[htbp]
\vbox{\hfil\epsfbox{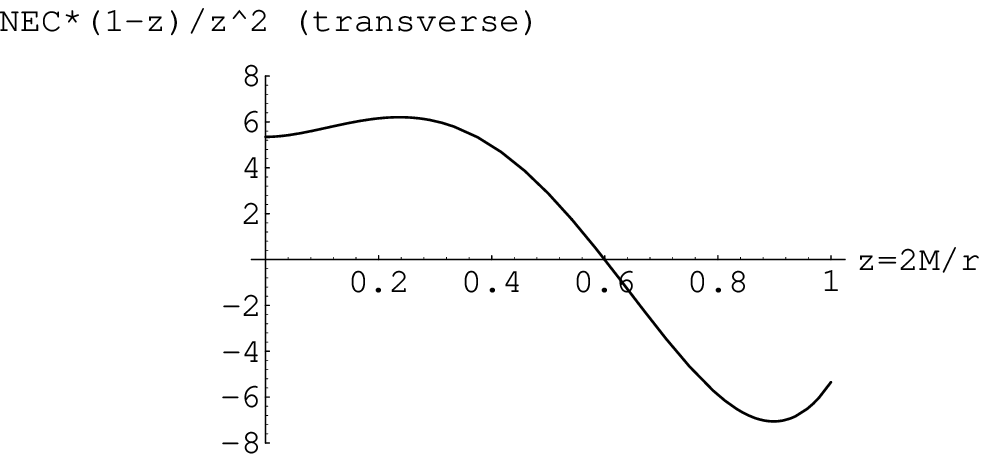}\hfil}
\caption[NEC (transverse)]
{\label{fig10}
$\rho+p$: This graph shows the NEC (evaluated using the semi-analytic
model, multiplied by $(1-z)$ to control the singularity at the
horizon, and divided by $z^2$ to suppress the asymptotic behaviour
at infinity) on transverse null vectors as a function of $z=2M/r$.
Note that the curve changes sign near $z=0.6$.  The NEC is {\em
satisfied} on transverse null vectors far away from the black hole,
and {\em violated} sufficiently near the black hole. (In particular
the NEC is {\em violated} on the unstable photon orbit at $r=3M$.)}
\end{figure}

The fact that $\rho-\tau-2f$ is negative over the entire range
$z\in[0,1]$, is enough to imply that at least in the outgoing radial
direction the NEC is violated throughout the region exterior to
the event horizon. {\em Ipso facto} all the pointwise energy
conditions (null, weak, strong, and dominant) are violated throughout
the entire region outside the event horizon, all the way to spatial
infinity.

The pointwise energy conditions are violated in the sense that at
every point outside the event horizon there is at least one null
or timelike vector violating the conditions.  It is certainly not
true that all null or timelike vectors violate the energy conditions,
nor is it true that the energy condition violations are large. What
is true is that the violations are widespread.

Furthermore, this implies that the averaged null energy condition
(ANEC) is violated on outgoing radial null geodesics. (More precisely,
the one-sided ANEC that is cut off at the event horizon is violated).
This automatically implies violations of (one-sided) averaged weak
(AWEC) and averaged strong (ASEC) energy conditions.

One can also immediately see that the null energy condition is
violated along the unstable photon orbit at $r=3M$ ($z=2/3$).

We could now add more detailed analyses of exactly which energy
condition is violated where, along the lines reference~\cite{Visser96a},
but the limited additional insight to be gained does not seem to
warrant it.

\subsection{Some analytic results}

In the outgoing null direction we have the exact result [{\em cf.}
eq. (\ref{E-nec-preliminary})]

\begin{eqnarray}
\langle U | T^{\hat\mu\hat\nu} | U \rangle  \; 
\ell^+_{\hat\mu} \ell^+_{\hat \nu} 
&=&   
\rho -\tau - 2 f, 
\nonumber\\
&=&
-4 {z^2\over1-z} 
\left[ 
{F(z)-F(0)\over2} + \xi \left( {z^4\over4} - {z^5\over5} \right) 
\right].
\end{eqnarray}

\noindent
But I have already shown  [{\em cf.} eq. (\ref{E-O3}) and the
discussion on p. \pageref{D-O3}]  that $F(z) = F(0) + O(z^3)$.
Therefore, as previously asserted

\begin{equation}
\langle U | T^{\hat\mu\hat\nu} | U \rangle  \; 
\ell^+_{\hat\mu} \ell^+_{\hat \nu} 
= O(z^5).
\end{equation}

\noindent
In fact, from eq. (\ref{E-diff-F}) we have $F(z) = F(0) + z^3
Q_1(z)$, with $Q_1(z)$ having a power series that starts off with
a constant $z^0$ term. Thus

\begin{equation}
\langle U | T^{\hat\mu\hat\nu} | U \rangle  \; 
\ell^+_{\hat\mu} \ell^+_{\hat \nu} 
= -4 {z^5\over1-z} 
\left[ 
{Q_1(z)\over2} + \xi \left( {z\over4} -{z^2\over5} \right) 
\right].
\end{equation}

\noindent
It is this explicit prefactor of $z^5/(1-z)$ that is responsible
for most of the structure as seen in figures \ref{fig7} and
\ref{fig8}.

On the other hand, in the ingoing null direction it is easy to
obtain the general result

\begin{eqnarray}
\langle U | T^{\hat\mu\hat\nu} | U \rangle  \; 
\ell^-_{\hat\mu} \ell^-_{\hat \nu} 
&=&   
\rho -\tau + 2 f, 
\nonumber\\
&=&
+ {\xi\over5}\; p_\infty z^2 (1+z+z^2+z^3-4z^4) 
- 2 {z^2\over1-z} F(z),
\\
&=&
+{\xi\over5}\; p_\infty \; z^2 (1-z)(1+2z+3z^2+4z^3) 
- 2 {z^2\over1-z} F(z).
\nonumber\\
&&
\end{eqnarray}

\noindent
But from eq. (\ref{E-factor-2}) we know that $F(z)$ contains an
explicit factor of $(1-z)^2$ so that $F(z) = (1-z)^2 Q_2(z)$ with
$Q_2(z)$ having a power series that starts off with a constant $z^0$
term. Thus

\begin{equation}
\langle U | T^{\hat\mu\hat\nu} | U \rangle  \; 
\ell^-_{\hat\mu} \ell^-_{\hat \nu} 
= z^2 (1-z) 
\left[ {\xi\over5}\; p_\infty (1+2z+3z^2+4z^3) - 2 Q_2(z) \right].
\end{equation}

\noindent
It is this explicit prefactor of $z^2(1-z)$ that is responsible
for most of the structure as seen in figure \ref{fig6}.

Finally, in the transverse direction one has the general result

\begin{eqnarray}
\langle U | T^{\hat\mu\hat\nu} | U \rangle  \; 
\ell^\perp_{\hat\mu} \ell^\perp_{\hat \nu} 
&=&   
\rho + p, 
\nonumber\\
&=&
-f_0 {z^2\over1-z} + {\xi\over10}\; p_\infty z^2 (1+z+z^2+z^3-9z^4) 
\nonumber\\
&& 
\qquad +2p(z) - 2 {z^2\over1-z} F(z),
\\
&=&
{z^2\over1-z} 
\Bigg\{ 
- f_0 +{\xi\over10}\; p_\infty \; (1-z)(1+2z+3z^2-9z^3) 
\nonumber\\
&& 
\qquad 
+2 (1-z) {p(z)\over z^2} - 2 F(z) \Bigg\}.
\end{eqnarray}

\noindent
But $p(z) = z^4 Q_3(z)$ where $Q_3(z)$ has a power series that
starts off with a constant $z^0$ term. Thus

\begin{eqnarray}
\langle U | T^{\hat\mu\hat\nu} | U \rangle  \; 
\ell^\perp_{\hat\mu} \ell^\perp_{\hat \nu} 
&=& {z^2 \over (1-z)} 
\Big[ 
-f_0 + {\xi\over10}\; p_\infty (1-10z^4-9z^5) 
\nonumber\\
&&\qquad 
+2 (1-z) z^2 Q_3(z) 
- 2 F(z) \Big].
\end{eqnarray}

\noindent
It is this explicit prefactor of $z^2/(1-z)$ that is responsible
for most of the structure as seen in figures \ref{fig9} and
\ref{fig10}.

\subsection{Some explicit results for the three-term model}

This generic behaviour can easily be checked analytically for the
simple three-term model developed in this paper.  A little brute
force yields

\begin{eqnarray}
\rho-\tau+2f &=&  p_\infty z^2 (1-z)
  \Bigg[ 4\left({\xi\over20} - {k_5\over24} - {k_6\over20}\right) + 
         \left({2\xi\over5} - {k_5\over3} - {2k_6\over5} \right) z 
\nonumber\\
&&\qquad 
       + \left({3\xi\over5} -{k_5\over2} - {3 k_6\over5} \right)z^2 
       + \left( {4\xi\over5} -{4 k_6\over5} \right)z^3 \Bigg].
\\
\rho-\tau-2f &=& p_\infty {z^5\over1-z}
\Bigg[  
{2k_5\over3} - \left(\xi+{k_5\over2}-k_6\right) z +
\left({4\xi\over5} - {4k_6\over5}\right)
\Bigg].
\\
\rho+p  &=&  p_\infty {z^2\over1-z}
  \Bigg[ \left({\xi\over20} - {k_5\over24} - {k_6\over20}\right) + 
         2k_4 z^2 +
         \left(-2k_4 - {7k_5\over3} \right) z^3 
\nonumber\\
&&\qquad 
       - \left(\xi +{9k_5\over4} - {5 k_6\over2} \right)z^4
       + \left( {9\xi\over10} -{12 k_6\over5} \right)z^5 \Bigg].
\end{eqnarray}

\noindent
The known analytic behaviour of the prefactors  in these expressions
serves as a consistency check on the numerical analysis used to
generate the figures.

\section{Discussion}

In summary, in this paper I have developed a systematic way of
building semi-analytic models for the stress-energy tensor in
Schwarzschild spacetime. For the Unruh vacuum I have carried the
program forward to the extent of explicitly deriving a three-parameter
approximation to the total stress-energy that successfully fits all
known data to better than $1\%$ accuracy. The model passes the
consistency test of correctly predicting the luminosity from the
fit to the transverse pressure.

In two appendices I sketch how this program can be extended to the
Hartle--Hawking and Boulware states.

A central result of this paper is the observation that violations
of the energy conditions, both pointwise and averaged, are ubiquitous
(though small) in the Unruh vacuum. This $(3+1)$--dimensional result
is qualitatively in agreement with the $(1+1)$--dimensional analytic
model considered in~\cite{Visser96c}. Furthermore in view of the
results quoted in~\cite{Visser96a,Visser96b} we know that this is
not a peculiarity of the Unruh vacuum, but that energy condition
violations are also widespread in the Hartle--Hawking and Boulware
vacuum states.

Note that I am claiming that the violations are widespread---I am
not claiming they are large. These are intrinsically quantum
mechanical effects with the typical scale of the effect near the
horizon being given by

\begin{equation}
\langle T^{\hat\mu\hat\nu} \rangle \approx {\hbar c^9\over (GM)^4}.
\end{equation}

Ford and Roman have argued~\cite{Ford-Roman94,Ford-Roman95a,Ford-Roman95b}
that in many situations the quantum inequalities may be of more
interest than the energy conditions themselves. It seems that even
if the energy condition violations are widespread, the quantum
inequalities may more stringently constrain the
dynamics~\cite{Ford-Roman95b}.  The semi-analytic model developed
in this paper may be of some interest in explicitly testing the
quantum inequalities. (The analysis of this paper is fully consistent
with qualitative features of the Ford-Roman results
of~\cite{Ford-Roman95a}, but the semi-analytic model of this paper
will allow one to evaluate the relevant integrals more accurately.
Note that although the coefficients were fitted using data from
outside the event horizon there is nothing to stop us from taking
the resulting model and applying it inside the event horizon.)

Finally, I remind the reader that issues of quantum mechanical
violation of the energy conditions are of central importance to
any attempt at taking the classical collapse theorems~\cite{Hawking-Ellis},
the classical topological censorship theorem~\cite{Topological-censorship},
or the classical positive mass theorem~\cite{Penrose-Sorkin-Woolgar}
and trying to see whether or to what extent these classical theorems
can be generalized into the semiclassical world.

\clearpage
\appendix
\section{Model building: Hartle--Hawking vacuum}

There is nothing particularly sacred about the Unruh vacuum when it
comes to model-building of the type discussed in this paper. Similar
procedures can be applied in the Hartle--Hawking vacuum as well. This
sort of modeling allows us to get a better handle on the underlying
rationale behind the Page approximation---this approximation is in
some sense (to be described below) the minimal form of the
stress-energy tensor compatible with regularity on the horizon and a
``thermal-bath plus curvature-corrections'' ansatz at spatial infinity.

\subsection{General analysis}

For general background information see~\cite{Fulling77}.  First,
since the Hartle--Hawking vacuum is to be regular on the both the
future horizon and past horizon we must have $f_+ \equiv 0 \equiv
f_-$. (And hence the flux is identically zero: $f(z)\equiv 0$.) At
asymptotic spatial infinity we want the stress-energy to look like
that of an thermal bath of radiation at the Hawking
temperature~\cite{Fulling77}.  That is:  we need to have $p(z) \to
p_\infty$ asymptotically as $z \to 0$, while $\rho(z)\to3p_\infty$,
and $\tau(z) \to -p_\infty$. (For higher-spin fields $\rho(z)\to3
g p_\infty$, $\tau(z) \to - g p_\infty$, and $p(z) \to +g p_\infty$,
with $g$ being the appropriate statistical weight.) 

Substituting the result for $f_0$ back into the general expression
for the stress-energy, the various components are seen to be

\begin{eqnarray}
\rho(z) &=& 
+{\xi\over10} \; p_\infty \; z^2 
\left( 1+z+z^2+z^3-9z^4 \right) + p(z) - {z^2\over1-z} F(z).
\\
\tau(z) &=&
-{\xi\over10} \; p_\infty \; z^2 
\left( 1+z+z^2+z^3+z^4 \right) + p(z) + {z^2\over1-z}  F(z).
\end{eqnarray}

\noindent
The transverse pressure is still well-behaved all the way down to the
horizon, so it still makes sense to look for a convergent
power-series.  But because of the known asymptotic behaviour, it is
more useful to introduce four new parameters ($k_0$,$k_1$,$k_2$,and
$k_3$; not present for the Unruh vacuum) and write:

\begin{equation}  
p(z) \equiv  p_\infty \; \sum_{n=0}^\infty k_n z^n . 
\end{equation}

\noindent
A brief calculation yields

\begin{eqnarray}
F(z) &=&  
+ p_\infty \Bigg\{
-2 k_{0}{(1-z)^2\over z^2} -3 k_{1} {1-z \over z}  
+ 2 k_{2} (1-z) - k_{3} {(1-z)^2\over2}
\nonumber\\
&& 
-(3k_{1}-2k_{2})\ln(z) 
\nonumber\\
&&
+ (1-z)
\left[ 
  \sum_{n=5}^\infty {(n-4) k_n\over(n-2)(n-1)}
  \left[ 1 + z + z^2 + \cdots + z^{n-3} - (n-2)z^{n-2} \right] 
\right] \Bigg\}.
\nonumber\\
&&
\end{eqnarray}

\noindent
Here I have kept the first few terms separate for clarity. Note that
$k_4$ does not contribute to $F(z)$.

\noindent
We are really only interested in the quantity

\begin{eqnarray}
{z^2\over1-z} F(z) 
&=&  
+ p_\infty \Bigg\{
{-2 k_{0} (1-z)} -3 k_{1} z + 2 k_{2} z^2 
- k_{3} {z^2(1-z)\over2} 
\nonumber\\
&&
-(3k_{1}-2k_{2}){z^2\ln(z)\over1-z} 
\nonumber\\
&&
+ z^2 
\left[ 
  \sum_{n=5}^\infty {(n-4) k_n\over(n-2)(n-1)}
  \left[ 1 + z + z^2 + \cdots + z^{n-3} - (n-2)z^{n-2} \right] 
\right] \Bigg\}
\nonumber\\
&=&  
+ p_\infty \Bigg\{
{-2 k_{0} (1-z)} -3 k_{1} z + 2 k_{2} z^2 
- k_{3} {z^2(1-z)\over2} 
\nonumber\\
&&
-(3k_{1}-2k_{2}){z^2\ln(z)\over1-z} 
+z^2 \left[ \sum_{n=5}^\infty {(n-4) k_n\over(n-2)(n-1)} \right]
\nonumber\\
&&
+ z^3 
\left[ 
  \sum_{n=5}^\infty {(n-4) k_n\over(n-2)(n-1)}
  \left[ 1 + z + z^2 + \cdots + z^{n-4} - (n-2)z^{n-3} \right] 
\right] \Bigg\}.
\nonumber\\
&&
\end{eqnarray}

\noindent
Which can now be substituted into the stress-energy to show

\begin{eqnarray}
\rho(z) &=& 
p_\infty 
\Bigg\{ 
+{\xi\over10} \; z^2  \left( 1+z+z^2+z^3-9z^4 \right)
\nonumber\\
&&\qquad
+k_{0} (3 -2z) + 4 k_{1} z -  k_{2} z^2 + k_{3} {z^2(1+z)\over2} 
\nonumber\\
&&\qquad
+ (3k_{1}-2k_{2}){z^2\ln(z)\over1-z}
-z^2 \left[ \sum_{n=5}^\infty {(n-4) k_n\over(n-2)(n-1)} \right]
+ \sum_{n=4}^\infty k_n z^n
\nonumber\\
&&\qquad
- z^3 \sum_{n=5}^\infty {(n-4) k_n\over(n-2)(n-1)}
  \left[ 1 + z + z^2 + \cdots + z^{n-4} - (n-2)z^{n-3} \right] \Bigg\},
\nonumber\\
&&\qquad
\\
\tau(z) &=&
p_\infty
\Bigg\{ 
-{\xi\over10}  \; z^2 \left(  1+z+z^2+z^3+z^4 \right)
\nonumber\\
&&\qquad
- k_{0}(1-2z) -2 k_{1} z +3 k_{2} z^2 - k_{3} {z^2(1-3z)\over2} 
\nonumber\\
&&\qquad
- (3k_{1}-2k_{2}){z^2\ln(z)\over1-z}
+z^2 \left[ \sum_{n=5}^\infty {(n-4) k_n\over(n-2)(n-1)} \right]
+ \sum_{n=4}^\infty k_n z^n
\nonumber\\
&&\qquad
+ z^3 \sum_{n=5}^\infty {(n-4) k_n\over(n-2)(n-1)}
  \left[ 1 + z + z^2 + \cdots + z^{n-4} - (n-2)z^{n-3} \right] \Bigg\}.
\nonumber\\
\end{eqnarray}

\noindent
From the preceding analysis it is clear that this model satisfies
all the known properties of the stress-energy tensor in the
Hartle--Hawking vacuum (anomalous trace, covariant conservation,
asymptotic behaviour both at spatial infinity and the horizon).
Consequently these equations provide a general formalism for the
stress-energy tensor in the Hartle--Hawking vacuum.

At first glance the presence of the logarithmic terms may be
disturbing. Note that the logarithms show up only in the combination
$S(z)\equiv [z^2\ln(z)]/(1-z)$. This combination remains finite
both at the horizon ($\lim_{z\to1} S(z) = -1$) and at spatial
infinity ($\lim_{z\to0}S(z) =0$), so the logarithmic terms should
not be excluded {\em a priori}.

There is however a popular ansatz that justifies eliminating the
logarithmic terms. Near spatial infinity we expect the stress-energy
tensor to be that of a red-shifted thermal bath of radiation with
curvature corrections. Schematically:

\begin{equation}
\langle H | T^{\hat\mu\hat\nu} | H \rangle(z) = 
{\langle H | T^{\hat\mu\hat\nu} | H \rangle|_\infty \over (1-z)^2} 
+\hbox{``curvature-corrections''}?
\end{equation}

Two versions of this ansatz will now be used to more precisely
fix the form of the stress tensor.

\subsection{Weak thermal bath ansatz}

\noindent
Since the curvature is proportional to $M/r^3$, we expect
the ``curvature corrections'' to be (at worst) of order $z^3$, in
which case we have the conservative ansatz:

\begin{equation}
\langle H | T^{\hat\mu\hat\nu} | H \rangle|_z = 
\langle H | T^{\hat\mu\hat\nu} | H \rangle|_\infty 
\left( 1 + 2 z + 3 z^2 \right) + O(z^3) ? 
\label{Ansatz1}
\end{equation}

\noindent
If we adopt this ansatz, it enforces very specific choices on the
first few coefficients. In fact, by looking at $p(z)$ it is trivial to
see that $k_{0} = 1g$, and $k_{1} = 2g$, while $k_{2} = 3g$. (Here $g$
is the statistical weight, $g=1$ for scalars. Also, remember to use
the appropriate value of $\xi$ for higher spin.) This is enough to
make the coefficients in front of the logarithmic terms vanish.  It is
easy to see that the $O(1)$ pieces of $\rho(z)$ and $\tau(z)$ are $3g$
and $-g$ respectively, while the $O(z)$ terms are $6g$ and $-2g$
respectively.  Additionally, the $O(z^2)$ pieces of $\rho(z)$ and
$\tau(z)$ are proportional to

\begin{equation} 
{\xi\over10} - 3g + {k_{3}\over2} 
- \sum_{n=5}^\infty {(n-4) k_n\over(n-2)(n-1)} 
= 3\times3g?
\end{equation}

\noindent
and

\begin{equation}
-{\xi\over10} + 9g - {k_{3}\over2} 
+ \sum_{n=5}^\infty {(n-4) k_n\over(n-2)(n-1)} = -3g?
\end{equation}

\noindent
respectively. So the conservative ansatz (\ref{Ansatz1}) can be
simultaneously satisfied for {\em all} components of the stress-energy
provided we pick
\begin{equation}
k_{3} = 
2 \left[ 
12g - {\xi\over10} + \sum_{n=5}^\infty {(n-4) k_n\over(n-2)(n-1)} 
\right].
\end{equation}

\noindent
Note that with this choice of coefficients, not only do we satisfy
the ansatz given above, but we also guarantee that the stress-energy
tensor will agree with Page's analytic approximation~\cite{Page82,Visser96a}
at least to order $O(z^3)$. With this ansatz in place

\begin{equation}  
p(z) \equiv  p_\infty \; 
\left[  g(1 + 2 z + 3 z^2)  + k_{3} z^3+ \sum_{n=4}^\infty k_n z^{n} \right]. 
\end{equation}

\noindent
The other components of the stress-energy tensor reduce to

\begin{eqnarray}
\rho(z) &=& 
p_\infty 
\Bigg\{ 
+3 g(1 + 2 z + 3 z^2) + k_{3} {z^3\over2} + {\xi\over10} z^3 (1+z+z^2-9z^3)
+ \sum_{n=4}^\infty k_n z^n
\nonumber\\
&&\qquad
- z^3  \sum_{n=5}^\infty {(n-4) k_n\over(n-2)(n-1)}
  \left( 1 + z + z^2 + \cdots + z^{n-4} -(n-2)z^{n-3} \right) \Bigg\},
\nonumber\\
&&
\\
\tau(z) &=&
p_\infty
\Bigg\{ 
-g(1 + 2 z + 3 z^2) + k_{3} {3z^3\over2} -{\xi\over10} z^3 (1+z+z^2+z^3)
+ \sum_{n=4}^\infty k_n z^n
\nonumber\\
&&\qquad
+ z^3  \sum_{n=5}^\infty {(n-4) k_n\over(n-2)(n-1)}
  \left( 1 + z + z^2 + \cdots + z^{n-4} - (n+2)z^{n-3} \right) \Bigg\}.
\nonumber\\
&&
\end{eqnarray}

\noindent
This completes the weak ansatz. Note that $k_3$ is not a free variable
and that it is $k_4$ and higher order coefficients that uniquely fix
the stress tensor.

\subsection{Strong thermal bath ansatz}

A more radical ansatz is to assert that the ``curvature-corrections''
to the stress-energy tensor should be of asymptotic order
$(curvature)^2$, that is of asymptotic order $z^6$. In this case
we assert

\begin{equation}
\langle H | T^{\hat\mu\hat\nu} | H \rangle|_z = 
\langle H | T^{\hat\mu\hat\nu} | H \rangle|_\infty 
\left( 1 + 2 z + 3 z^2 + 4 z^3 + 5 z^4 +6 z^5 \right) + O(z^6) ? 
\label{Ansatz2}
\end{equation}

\noindent
We should note in particular that the Page approximation satisfies
this more radical ansatz. By considering $p(z)$ we automatically
deduce $k_0=1g$, $k_1=2g$, $k_2=3g$, $k_3=4g$, $k_4=5g$, and
$k_5=6g$, with $k_6$ and higher being left free by this ansatz.
Instead of using brute force it is useful to first consider a
subsidiary ansatz for $p(z)$:

\begin{equation}
p(z) = p_\infty \; g ( 1 + 2z + 3z^2 +4z^3 +5z^4 +6z^5 + k_6).
\end{equation}

\noindent
It is now straightforward to compute

\begin{eqnarray}
{z^2\over1-z} F(z) &=& p_\infty \; g
\Big[ 
(-2-4z+4z^2+2z^3) 
+ {1\over2} z^2 (1+z+z^2-3z^2) 
\nonumber\\
&&\qquad
+{k_6\over 10} z^2 (1+z+z^2+z^3-4z^3)
\Big].
\end{eqnarray}

\noindent
Inserting this into the stress-energy tensor

\begin{eqnarray}
\rho &=& \rho_\infty \; g \Big\{  
(1 + 2z + 3z^2 +4z^3 +5z^4 +6z^5 + k_6)
+(+2+4z-4z^2-2z^3) 
\nonumber\\
&&\qquad
- {1\over2} z^2 (1+z+z^2-3z^2) 
-{k_6\over 10} z^2 (1+z+z^2+z^3-4z^3)
\nonumber\\
&&\qquad
+{(\xi/g)\over10} z^2 (1+z+z^2+z^3-9z^2) \Big\}.
\end{eqnarray}

\noindent
Collecting terms 

\begin{eqnarray}
\rho &=& \rho_\infty \; g \Bigg\{  
3 
+ 6z 
+ z^2\left(-1+{(\xi/g)-k_6-5\over10}\right) 
+ z^3\left(+2+{(\xi/g)-k_6-5\over10}\right)
\nonumber\\
&&\qquad
+ z^4\left(+5+{(\xi/g)-k_6-5\over10}\right)
+ z^5\left(+8+{(\xi/g)-k_6-5\over10}\right)
\nonumber\\
&&\qquad
+ z^6\left({14k_6-9(\xi/g)\over10}\right)
\Bigg\}.
\end{eqnarray}

\noindent
This can be made to satisfy the strong thermal bath ansatz above if we
pick $(\xi/g)-k_6-5=100$, that is

\begin{equation}
k_6 = {\xi\over g} - 105.
\end{equation}

\noindent
Once we do this

\begin{eqnarray}
p &=& p_\infty g \left\{  
(1+2z+3z^2+4z^3+5z^4+6z^5) + \left({\xi\over g}-105\right)
\right\}.
\\
\rho &=& p_\infty g \left\{  
3(1+2z+3z^2+4z^3+5z^4+6z^5) + \left({\xi\over2g}-147\right)
\right\}.
\\
\tau &=& p_\infty g \left\{  
-(1+2z+3z^2+4z^3+5z^4+6z^5) + \left({\xi\over2g}-63\right)
\right\}.
\end{eqnarray}

\noindent
(The computation for $\tau$ is completely analogous to that just
performed for $\rho$.) This is exactly Page's analytic
approximation~\cite{Page82,Visser96a}.

What does this computation tell us? If we take the strong ansatz,
(that the stress-energy is a red-shifted thermal bath with curvature
corrections of order $z^6$), and supplement this with the requirement
that there be no terms of order higher than $z^6$, the we are led
uniquely to the Page approximation for the stress-energy tensor
surrounding a black hole. That is, the Page approximation is (in the
sense described above) the minimal ansatz compatible with general
properties of the stress-energy tensor.

Of course the Page approximation is not exact, and the stress-energy
around Schwarzschild black holes does have some higher order
contributions. Can we say anything about $O(z^7)$ terms and higher?
Indeed yes, when adding $O(z^7)$ terms and higher to $p(z)$ we should
be careful to not destroy the thermal bath ansatz for lower order
terms. Suppose we write

\begin{equation}
p(z) = p_{Page}(z) + \delta p(z).
\end{equation}

\noindent
Then we have

\begin{eqnarray}
\rho(z) &=& \rho_{Page}(z) - {z^2\over1-z} \delta F(z).
\\
\tau(z) &=& \tau_{Page}(z) + {z^2\over1-z} \delta F(z).
\end{eqnarray}

\noindent
So as to not destroy the strong thermal bath ansatz we need both
$\delta p(z)=O(z^6)$ or higher and $\delta F(z) = O(z^4)$ or
higher. We have already seen that we cannot possibly achieve this with
any monomial in $z$. The best we can hope for is to find some suitable
binomial such as 

\begin{equation}
\delta p_n(z) = z^n -\epsilon_n z^{n+1}. 
\end{equation}

\noindent
In which case 

\begin{eqnarray}
\delta F_n(z) &=& 
(1-z) \Bigg\{  
{(n-4)\over(n-2)(n-1)} 
     \left[ 1 + z + z^2 + \cdots + z^{n-3} - (n-2)z^{n-2} \right] 
\nonumber\\
&&\qquad
-\epsilon_n  
{(n-3)\over(n-1)n} 
     \left[ 1 + z + z^2 + \cdots + z^{n-2} - (n-1)z^{n-1} \right]
\Bigg\}
\nonumber\\
&&\qquad
\end{eqnarray}

\noindent
So if we pick $\epsilon_n = [n(n-4)]/[(n-2)(n-3)]$ there are massive
cancellations and

\begin{equation}
\delta F_n(z) =  -{n-4\over n-2} z^{n-2} (1-z)^2.
\end{equation}

\noindent
That is to say

\begin{equation}
{z^2\over1-z}\delta F_n(z) =  -{n-4\over n-2} z^n (1-z).
\end{equation}

\noindent
Consequently the binomials

\begin{equation}
\delta p(z) = z^n -{n(n-4)\over(n-2)(n-3)} z^{n+1}
\end{equation}

\noindent
(for $n\geq6$) are a useful ``basis'' for the pieces of the
stress-energy tensor that go beyond the Page approximation. These
basis elements are useful in the sense that they do not perturb the
lower order pieces of the stress-energy.

I shall report elsewhere the results of performing such fits to
the Anderson--Hiscock--Samuel data.

\clearpage
\section{Model building: Boulware vacuum}

Unsurprisingly, a similar analysis can be applied in the Boulware
vacuum. Again, for general background information see~\cite{Fulling77}.
First, since there is no net flow of radiation in the Boulware
vacuum we must have $f_+\equiv f_0\equiv f_-$. (And hence $f(z)=0$).
At asymptotic spatial infinity we want the stress-energy to be as
small as possible. Since if nothing else the anomaly will generate
terms of order $O(z^6)$, Christensen and Fulling were led to
tentatively suggest $\langle H | T^{\hat\mu\hat\nu} | H \rangle =
O(z^6)$~\cite[page 2096]{Fulling77}.

To at least force the terms of order $O(z^2)$ to vanish we need
to enforce

\begin{equation}
{\xi\over10} + G(0) + 2 f_0 = 0.
\end{equation}

\noindent
That is  

\begin{equation}
f_0 = -{\xi\over20} \; p_\infty - {G(0)\over2}.
\end{equation}

\noindent
(This is formally very similar to what we found for the Unruh
vacuum, up to a few critical minus signs.) Substituting this result for
$f_0$ back into the general expression for the stress-energy, the
various components are seen to be

\begin{eqnarray}
\rho(z) &=& 
-{\xi\over10} \; p_\infty \; z^6 {10-9z\over1-z}
+ 2 p(z)
+ {z^2\over1-z} \left[G(z)-G(0)\right] .
\\
\tau(z) &=&
+{\xi\over10} \; p_\infty \; {z^7\over1-z}
- {z^2\over1-z}  \left[ G(z) - G(0) \right] .
\end{eqnarray}

I have actually pulled a minor swindle to get to this point because
$G(0)$, $G(z)$, and $f_0$ are all ill-defined infinite quantities
in the Boulware vacuum. This happens because $p(z)$ diverges at
the horizon, and the integral used to define $G(z)$ does not
converge. Fortunately {\em this does not matter}, since the final
expression for the stress-energy contains only terms of the type

\begin{equation}
\bar G(z) \equiv \left[ G(z) - G(0) \right] \equiv 
- \int_0^z \left({2\over {\bar z}^3}-{3\over {\bar z}^2} \right) 
            \; p(\bar z) \; d{\bar z}.   
\end{equation}

\noindent
This integral converges provided $p(z)$ is of order $O(z^3)$ or smaller
at spatial infinity. The divergence at the horizon does not matter
because the range of integration does not include the horizon.

We can further define

\begin{equation}
\bar F(z) \equiv \left[ F(z) - F(0) \right] \equiv
-\int_0^z {\bar z}^2 \; (1-{\bar z}) 
\; {d\over d\bar z}\left({p(\bar z)\over{\bar z}^4}\right) \; d{\bar z}.
\end{equation}

\noindent
Then, integrating by parts, it is easy to show that 

\begin{equation}
\bar G(z) = -{1-z\over z^2} p(z)  - \bar F(z).
\end{equation}

\noindent
Doing this requires only the mild assumption that $p(z)$ is is of
order $O(z^3)$ or smaller at spatial infinity, which we already
had to assume anyway.

The components of the stress-energy tensor can now be rewritten as

\begin{eqnarray}
\rho(z) &=& 
-{\xi\over10} \; p_\infty \; z^6 {10-9z\over1-z} 
+ p(z) - {z^2\over1-z} \bar F(z).
\\
\tau(z) &=&
+{\xi\over10} \; p_\infty \; {z^7\over1-z} 
+ p(z) + {z^2\over1-z}  \bar F(z).
\end{eqnarray}

\noindent
The transverse pressure is still well-behaved {\em until} you get
to the horizon, so it still makes sense to look for a power-series;
though now we expect this power series to have radius of convergence
of one, and to diverge at the horizon.  Because of the minimal $O(z^3)$
asymptotic behaviour I have argued for above, it is instructive to write

\begin{equation}  
p(z) \equiv  p_\infty \; 
\left[ k_{3} z^3+ \sum_{n=4}^\infty k_n z^{n} \right]. 
\end{equation}

\noindent
A brief calculation yields

\begin{equation}
\bar F(z) =  
+ p_\infty \left\{
 k_{3} \left(z-{z^2\over2}\right)  + 
\left[ 
  \sum_{n=5}^\infty (n-4) k_n
  \left( {z^{n-1}\over n-1} - {z^{n-2}\over n-2}\right) 
\right] \right\}
\end{equation}

\noindent
Which can now be substituted into the stress-energy to show 

\begin{eqnarray}
\rho(z) &=& 
p_\infty 
\Bigg\{ -{\xi\over10} \; z^6 {10-9z\over1-z} 
-  k_{3} {z^4\over2(1-z)} + \sum_{n=4}^\infty k_n z^n
\nonumber\\
&&\qquad
- {z^2\over1-z} \sum_{n=5}^\infty (n-4) k_n
  \left( {z^{n-1}\over n-1} - {z^{n-2}\over n-2} \right) \Bigg\},
\\
\tau(z) &=&
p_\infty
\Bigg\{ +{\xi\over10}  \;  {z^7\over1-z}
+  k_{3} {z^3(4-3z)\over1-z} + \sum_{n=4}^\infty k_n z^n
\nonumber\\
&&\qquad
+ {z^2\over1-z} \sum_{n=5}^\infty (n-4) k_n
  \left( {z^{n-1}\over n-1} - {z^{n-2}\over n-2} \right) \Bigg\}.
\end{eqnarray}

\noindent
We again see from the preceding analysis that this model satisfies
all the known properties of the stress-energy tensor in the Boulware
vacuum (anomalous trace, covariant conservation, asymptotic behaviour
both at spatial infinity and the horizon). Consequently these
equations provide a general formalism for the stress-energy tensor
in the Boulware vacuum.

It is possible to improve the situation by making several simplifying
ansatze.  If we assume that all non-zero components of the
stress-energy tensor should asymptotically be of the same order,
then we must set $k_{3}=0$. If (following Christensen--Fulling) we
make the stronger ansatz that that all non-zero components are of
order $O(z^6)$ one has $k_{3}=k_4=k_5=0$. In this case it is easy
to check that

\begin{eqnarray}
\rho(z) &=& p_\infty \; \left(-\xi + {3\over2} \; k_2 \right) z^6 + O(z^7).
\\
\tau(z) &=& p_\infty  \; {1\over 2} \; k_2 \; z^6 + O(z^7).
\\
p(z) &=& p_\infty \; k_2 \; z^6 + O(z^7).
\end{eqnarray}

\noindent
For $k_6 = -16$ this is compatible with the Brown--Ottewill analytic
approximation~\cite{Brown-Ottewill}. I have not found any nice
principle that would uniquely lead from this general decomposition
to the full Page--Brown--Ottewill approximation.

Serious model building is now simply a matter of getting good enough
data to make reasonable estimates of the various coefficients $k_n$.
One might even hope that with a little more work it might prove
easier to analytically calculate these coefficients in
preference to the numerically intensive work required to calculate
the stress-energy components directly.

\section*{Acknowledgements}

{}

I wish to thank Bruce Jensen and Adrian Ottewill for kindly making
available machine-readable tables of the Unruh vacuum numeric data
used in this analysis.

I wish to thank Paul Anderson for kindly making available
machine-readable tables of the Hartle--Hawking vacuum numeric data used
in this analysis.

I also wish to thank Nils Andersson, Paul Anderson, \'Eanna Flanagan,
Larry Ford, Eric Poisson, and Tom Roman for their comments.

The numerical analysis in this paper was carried out with the aid
of the Mathematica symbolic manipulation package.

This research was supported by the U.S. Department of Energy.



\begin{thebibliography}{99}
\bibitem{JLO-consistency}
In fact there are two minor problems with the scalar data of
reference~\cite{JLO91}. In figure 4 on page 4143 the difference
data for the radial tension ($\tau_U-\tau_H$) is correctly plotted
but has inadvertently been multiplied by an extraneous factor of
one-half before being added to Howard's estimate for the radial
tension. That is: $\tau_{plotted} = \tau_{H} + {1\over2}(\tau_U-\tau_H)$,
or $\tau_U = 2 \,\tau_{plotted} - \tau_H$. A second minor problem
with the energy density is discussed below.
\bibitem{JLO-energy-density}
In fact the semi-analytic model correctly reproduces the
Jensen--McLaughlin--Ottewill data for the radial tension (once
allowance for the extraneous factor of one-half has been made).
The semi-analytic fit also reproduces the data for the energy
density in the range $r\in[2M,4M]$, but exhibits small but systematic
deviations (too large to simply be numerical slop) in the range
$r\in[4M,5M]$. I attribute these deviations to a bug in the
integration routine used by Jensen--McLaughlin--Ottewill to calculate
their density data from their pressure data. As a check on this I
have evaluated the anomalous trace using the Jensen--McLaughlin--Ottewill
data and find that it in the range $r\in[4M,5M]$ it does not equal
$96 \, p_\infty \,z^6$, the difference being completely ascribable
to this bug in their calculation of the energy density.
\end{thebibliography}
\end{document}